\documentclass[12pt]{article}

\def\be{\begin{equation}}
\def\ee{\end{equation}}
\def\bea{\begin{eqnarray}}
\def\eea{\end{eqnarray}}

\usepackage{amssymb,latexsym}
\usepackage{graphicx}
\usepackage{amssymb, amsmath, amsopn, amsthm} 
\usepackage[section]{placeins}

\makeatletter \@addtoreset{equation}{section} 
\makeatother

\begin{document}
\begin{titlepage}
\thispagestyle{empty} 
\begin{flushright}
    SU-ITP-2006-31\\
November 16, 2006
\end{flushright}

\vspace{30pt} 

\begin{center}
    { \LARGE{\bf O'KKLT}}
    
    \vspace{40pt}
    
  {\large  
    Renata Kallosh and Andrei Linde}
    
    \vspace{10pt}

    \vspace{10pt} { \ Department of Physics,
    Stanford University, Stanford, CA 94305}
    
    \vspace{20pt}
 \end{center}

\begin{abstract}
We propose  to combine  the quantum corrected O'Raifeartaigh model, which has a dS minimum near the origin of the moduli space,  with the KKLT model with an  AdS minimum. The combined  effective N=1 supergravity model, which we call O'KKLT, has a dS minimum with all moduli stabilized. Gravitino in the O'KKLT model tends to be light in the regime of validity of our approximations. We show how one can construct models with a light gravitino and a high barrier protecting vacuum stability during the cosmological evolution. 
\end{abstract}

\vspace{10pt} 
\end{titlepage}

\newpage   \baselineskip 5.5 mm

\section{Introduction}

The O'Raifeartaigh model \cite{O'Raifeartaigh:1975pr}  provides the  simplest mechanism of  spontaneous breaking of global supersymmetry. Quantum  field theory corrections in this model lead to a stable minimum with positive energy at the origin of the moduli space. This model was studied and developed in various directions, 
see e.g. \cite{Huq:1975ue,Witten:1981kv,Poppitz:1998vd,Chacko:1998si,Intriligator:2006dd,Dine:2006ii,
Kitano:2006wz,Craig:2006kx}.

On the other hand, the KKLT scenario \cite{Kachru:2003aw} 
describes stabilization of  closed string theory moduli associated with no-scale supergravity. In this way the effective d=4  N=1 supergravity can be related to the superstring theory in the critical dimension d=10 where the stabilized modulus represents a volume of the compactified space.

One of the features  of KKLT construction is the uplifting of the AdS minimum to a dS minimum, which was originally achieved by the introduction of the ${\overline {D3}}$ brane. It was also developed as a  D-term uplifting \cite{Burgess:2003ic}, F-term uplifting \cite{Saltman:2004sn,Gomez-Reino:2006dk, Lebedev:2006qq,Dine:2006ii,Brummer:2006dg,Dudas:2006gr,Abe:2006xp}, K\"ahler/$\alpha'$ uplifting \cite{Westphal:2005yz}, etc.

Here we  propose to perform the uplifting by combining the two basic models,  O'Raifeartaigh and KKLT, into one model, which we will call O'KKLT.   We believe that this simple model  can serve as a prototype of the F-term uplifting with dynamical supersymmetry breaking in the KKLT scenario. This is in agreement with the observation in \cite{Chacko:1998si} that models with dynamical supersymmetry breaking near the origin of the moduli space reduce to the O'Raifeartaigh model, which has a stable minimum at the origin. Models of dynamical supersymmetry breaking include  simple models like \cite{Izawa:1996pk}
 as well as more realistic particle physics models with dynamical soft terms \cite{Dimopoulos:1997ww}. The fields of these  models  in principle may originate from the open string sector in models with intersecting branes in string theory.

Related models of uplifting  were already studied in \cite{Dudas:2006gr,Abe:2006xp} where  the ISS model \cite{Intriligator:2006dd} and other models with dynamical supersymmetry  breaking  where combined with the supergravity KKLT construction. For early attempts to do so see  \cite{Dine:2004is}. 
These models, as well as the one we propose, are realizing the general  scenario for the F-term uplifting  suggested in \cite{Gomez-Reino:2006dk}.  It was argued there that it is possible to construct a wide class of F-term uplifting models by adding to any
 sector $A$,  strongly influenced
by gravity, another sector $B$,  which separately  breaks supersymmetry in the rigid limit. Assuming  that
these two sectors do not directly mix and all dimensionful quantities in $B$ are small comparative to the Planck scale,  it was shown in \cite{Gomez-Reino:2006dk}
that the net effect of the sector $B$ is to provide an uplifting potential for
the sector $A$. 
With  $K = K_A + K_B$ and $W = W_A + W_B$, the uplifting potential has the form $V_{up} \approx e^{K} V_B$,
where $V_B$ is the potential of the sector $B$. In O'KKLT model, which we are going to develop here, the relevant uplifting term will be $\sim e^{K_{{KKLT}}}V_{O'}$, where $V_{O'}$ is the O'Raifeartaigh potential.

We will  address here the issue of the gravitino mass in the O'KKLT model. In the simplest KKLT-based models, the gravitino mass is directly related to the height of the barrier  stabilizing the volume modulus in string theory. This relation leads to a strong upper bound on the Hubble constant during inflation, $H \lesssim m_{3/2}$, which creates some tension between inflationary cosmology and the possibility of the low-scale SUSY breaking, to be studied on the LHC. A possible solution of this problem was proposed in \cite{Kallosh:2004yh}; here we will describe a generalization of this proposal for the case of the O'Raifeartaigh uplifted KKLT scenario.

\section{ The basic O'KKLT model}

We combine the two models as follows
\be
W=W_{O'}+ W_{KKLT}\ ,  \qquad K=K_{O'}+ K_{KKLT} \ .
\label{totalWK}\ee
Here the classic O'Raifeartaigh model $m\phi_1 \phi_2 + \lambda S \phi_1^2 - \mu^2 S$ will be used in the form where the relatively heavy ``O'Raifeartons'' $\phi_1, \phi_2$ are integrated out. We will assume that  all masses are much smaller than the Planck mass, i.e. $m, \mu \ll 1$  in units $M_{Pl}=1$. We will also assume that $ m^2 \gg  \lambda \mu^2$, to make sure that the fields $\phi_1$, $\phi_2$ vanish at the minimum of the potential. This condition  also leads to a simplification of the expression for  quantum corrections. In this case,
quantum corrections to the potential of the Coleman-Weinberg type  can be interpreted as quantum corrections to the K\"ahler potential of                                                                                                                                                                                                                                                                                                                                                                                                                                 $S$ field. Therefore we take
\be
W_{O'}=- \mu^2 S, \qquad K_{O'}= S\bar S - {(S\bar S)^2\over \Lambda^2}\,  
\label{model}\ee
 This form of the quantum corrected K\"ahler potential comes from the expansion of the one-loop expression $K^1_{O'}= S\bar S \left(1- {c\lambda^2\over 16 \pi^2} \log(1+ {\lambda^2 S\bar S\over m^2})\right)$ for ${\lambda^2 S\bar S\over m^2} \ll 1$. Here $\Lambda^{2}= {16\pi^{2}m^{2}\over c\lambda^{4}}$, $c = O(1)$, and we assume that $ {\lambda^{2} \over 16 \pi^2} \ll 1$ and $\Lambda^{2} \ll 1$. Our ``macroscopic'' model presented in eqs.  (\ref{model}) is valid for small $S$, when   
${S\bar S}\ll { m^2\over \lambda^2 }$.  The combination of  all of the required conditions  can be written as follows: 
\be\label{cond}
m ,~\mu ,~{\lambda \over 4\pi } \ll 1, \qquad  m \ll {\lambda^{2 }\over 4\pi}, \qquad \lambda \mu^2 \ll m^{2} , \qquad {S\bar S}\ll { m^2\over \lambda^2 } \ .
\ee

The supergravity potential $V_{O'}$ based on (\ref{model}) is given by
\be
V_{O'}(S, \bar S) = e^{K_{O'}}(|D_S W_{O'}|^2-3|W_{O'}|^2)\ ,
\ee
which leads to
\be
V_{O'}(S, \bar S) = \mu^4 e^{{S\bar S(\Lambda^2-S\bar S)\over \Lambda^2}} \left [{\left(\Lambda^2(1+(S\bar S)-2(S\bar S)^2\right)^2\over \Lambda^4- 4 \Lambda^2 S\bar S }- 3S\bar S\right] \ .
\ee 
The potential has a de Sitter minimum at $S=0$, at the origin of the moduli space. The value of the supergravity potential at the minimum is positive and the scale is defined by the parameter of the linear term in the superpotential. Near the minimum we find
\be
V_{O'}\approx  \mu^4 + {4  \mu^4 \over \Lambda^2}S\Bar S +...
\label{ORA}\ee
Thus, this model in the region of small $S$ looks like a good candidate for the uplifting term in the KKLT model.

Now we will combine it with the original  KKLT model \cite{Kachru:2003aw} with 
\be
W = W_0 + Ae^{-a\rho}\ , \qquad K = - 3 \ln[(\rho + \overline{\rho})  \ .
\label{simple}
\ee

The resulting O'KKLT model is given by                                                                                                                                                                                                                                                                                                                                                                                                               \be
W=W_0 + Ae^{-a\rho}- \mu^2 S, \qquad K= - 3 \ln(\rho + \overline{\rho})+ S\bar S - {(S\bar S)^2\over \Lambda^2} \ .
\label{model1}\ee
The complete potential $V(\sigma, \alpha, x, y)$ as a function of 4 scalars,
\be
\rho= \sigma+i\alpha \ , \qquad S=x+iy \ .
\ee
is easily computable in Mathematica, using e.g. the code in \cite{Kallosh:2004rs}, which was designed for the calculation of the N=1 supergravity potential for any number of chiral fields and arbitrary K\"ahler potential  and superpotential. However, we will be interested only in
 the region of small $S\bar S$, which is responsible for the uplifting. In this case the total potential can be represented  in a rather compact form
 \be
V_{O'KKLT}= V_{KKLT}(\rho, \bar \rho) + {V_{O'}(S, \bar S) \over (\rho+\bar \rho)^3} - i(S-\bar S) V_3  + (S+ \bar S) V_4 +  S\bar S V_5  \ .
\label{potential}\ee
Here
$V_3 (\rho, \bar \rho, S, \bar S) $,  $ V_4 (\rho, \bar \rho, S, \bar S)$ and   $ V_5 (\rho, \bar \rho, S, \bar S)$ depend on $S$, $\bar S$  polynomially.

Note that   the KKLT potential $V_{KKLT}(\rho, \bar \rho)$, taken separately, has an AdS minimum at the vanishing axion,  $\alpha = 0$,  and at some (large) value of $\sigma$. Here we consider the case when $a A W_0$ is negative and take into account that all dependence on axion in $V_{KKLT}(\rho, \bar \rho)$ is  in the term 
\be
 {a A W_0 e^{-a\sigma} \cos(a\alpha)\over 2 \sigma^2}
\label{cos} \ee 
The potential of the quantum corrected O'Raifeartaigh model $V_{O'}(S, \bar S)$, taken separately, has a minimum at $S=x+iy=0$. 

One could expect that the position of the minimum of the potential of the combined model may not differ much from the position of the  minimum of the two models taken independently. We should check whether this is indeed the case, to make sure that after the unification the values of all fields remain within the domain of validity of our approximation, which requires that ${S\bar S}\ll { m^2\over \lambda^2 }$. Our numerical examples support this expectation and show that the values of the axion fields $\alpha$ and $y$ in the minimum of the combined potential remain equal to zero, whereas the values of $\sigma$ and $x$ are only slightly shifted,  and the condition ${S\bar S}\ll { m^2\over \lambda^2 }$ is indeed valid for certain values of the parameters of our model. 

To understand this result analytically, we note that the  3d term in  (\ref{potential}) depends on the O'Raifeartaigh axion $y$ linearly and on the KKLT axion $\alpha$ via  $\sin(a\alpha)$. It  has the following structure:
\be
- i(S-\bar S)\, V_3 (\rho, \bar \rho, S, \bar S)= y \sin (a\alpha)~  {A \mu^2 e^{-a\sigma}(1+a\sigma)\over 4\sigma^3}  + {\cal O}(x^2, y^2)\ .
\ee 
The 4th term in (\ref{potential}) consists of 2 terms. The first one does not depend on  $\alpha$,  whereas the second one depends on $\cos (a\alpha)$,  but it is significantly smaller than the KKLT $\alpha$-dependent term shown in eq. (\ref{cos}). There are no other terms linear in $y$ in the potential. All other terms are either quadratic or higher power in $y^2$.
Therefore the linearized equation for $y$ and exact equation for $\alpha$ extremizing the potential (at fixed values of $\sigma$ and $x$) take the following form
\be
{\partial V\over \partial \alpha} = c_1 \sin (a\alpha) + c_2 y \cos (a\alpha)\ , \qquad{\partial V\over \partial y}= c_3 \sin (a\alpha) + c_4 y \ . 
\ee
This shows that the point $\alpha = y = 0$ remains the extremum of the potential after the uplifting. One can also show that it remains a minimum.

The potential at $\alpha=y=0$ is significantly simplified. One can now proceed with  evaluation of the change of its minimum with respect to $\sigma$ and $x$. Since the KKLT potential is uplifted at least via the second term in (\ref{potential}) we know that $\sigma$ changes a bit, as in the original KKLT model. This shift is small, exactly as in the original KKLT model where the uplifting occurs due to the ${\overline {D3}}$ brane.

The situation with the field $x$, the  real part of $S$, is the following. One can expand the potential in powers of $x$. The coefficients in such expansion are complicated functions of $\sigma$. However,  in the first approximation, we can  calculate the values of the linear and quadratic terms in $x$ at $\sigma=\sigma_{0} $ where $\sigma_{0} $ is the value of $\sigma$ at the AdS minimum before the uplifting. At this point we find that 
\be
V \approx c_0(\sigma_{0}) +  c_1 (\sigma_{0}) x - {c_2(\sigma_{0}) \over 2} x^2+...
\ee
According to (\ref{ORA}), (\ref{potential}), the O'Raifeartaigh model uplifts the AdS minimum with the depth $|V_{AdS}|  \approx 3 m^{2}_{3/2}$  \cite{Kallosh:2004yh} by ${\mu^{4}\over (2\sigma_{0})^{3}}$, so that ${\mu^{4}\over (2\sigma_{0})^{3}}+V_{AdS} = c_0(\sigma_{0}) \sim 10^{{-120}} \approx 0$. This implies that 
\be
m^{2}_{3/2} \approx {\mu^{4}\over 3(2\sigma_{0})^{3}} \ .
\ee

The value of the field $x$ in the uplifted minimum is determined by the condition $V'=0$, which gives
\be
x_{0}\approx {c_1(\sigma_{0})\over c_2(\sigma_{0})}  = {\sqrt 3 \Lambda^2 \over    6- \Lambda^2 }\approx {\sqrt 3  \over 6 }  \Lambda^2 \ .
\ee
In the derivation of this formula we used the total expression for the combined potential from Mathematica and the values of the $\sigma_{0}  $ from \cite{Kachru:2003aw}  in the form $a A e^{-a \sigma_{0}  }=m_{3/2}\sqrt {18 \sigma_{0}  }$. 

This result has an interesting and instructive interpretation in terms of a simpler model recently studied by Kitano  \cite{Kitano:2006wz}. He studied the supergravity model with 
\be
W_{K}=- \mu^2 S +C, \qquad K_{K}= S\bar S - {(S\bar S)^2\over \Lambda^2}\ .  
\label{Kitano}\ee
The difference with what we call quantum corrected O'Raifeartaigh model is the presence of the constant term $C$ in the superpotential. Without the term ${(S\bar S)^2\over \Lambda^2}$ in the K\"ahler potential  this would be the supergravity Polonyi model \cite{Polonyi:1977pj}, which is known to have a Minkowski vacuum at the fine-tuned Planckian value of the field $S$. Thus, the model (\ref{Kitano}) is a hybrid of the Polonyi model and the  quantum corrected O'Raifeartaigh model. 
It was shown by Kitano that   one can fine-tune the constant $C$ to get the Minkowski vacuum, as in the Polonyi model. However,  in this hybrid case the minimum of the potential appears at not at $S = O(1)$, as in the Polonyi model, but at $S\approx {\sqrt 3  \over 6 }  \Lambda^2 \ll 1$ \cite{Kitano:2006wz}.  
The total potential based on (\ref{Kitano})  at small $x$ and small $\Lambda^2$ can be represented, using \cite{Kallosh:2004rs}, in a compact form:
\be
V_{K}\approx \mu^4[ (1-3C^2) - 4 C x+ 2({2\over \Lambda^2}-C^2) x^2+...
\ee 
The field $x$ is stabilized at
$
x_{0}\approx {C \Lambda^2\over 2-C^2\Lambda^2}.
$
If $C^2$ is tuned to Minkowski vacuum, $C^2=1/3$, one  finds for small $\Lambda^2$  that $x_{0}\approx {\sqrt 3 \over 6-\Lambda^2}\Lambda^2 \approx {\sqrt 3 \over 6}\Lambda^2$, precisely as in the O'KKLT model. 

This result is pretty general; in particular, it is valid for the generalized KKLT models  to be discussed in the next section. The meaning of this result is that the KKLT model supplies the superpotential of the quantum corrected O'Raifeartaigh model  by the constant term (at fixed $\sigma = \sigma_{0}$).  In terms of Eqs. (\ref{Kitano}),  $W_{KKLT} (\sigma_{0})=C$  serves for adjusting the height of the minimum and making it the (nearly) Minkowski one.  In other words,  fine tuning of $W_{KKLT} (\sigma_{0})=C$ in the O'KKLT model allows us to achieve the cancellation  between the negative energy density in the AdS minimum of the KKLT and the positive energy density in the dS minimum of the quantum corrected O'Raifeartaigh model.

\begin{figure}[hbt]
 \centering 
\includegraphics[scale=0.9]{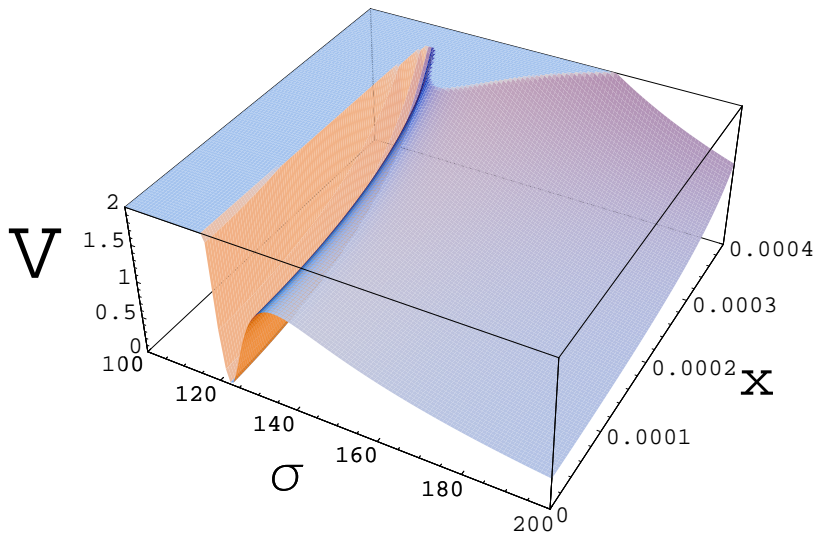} \caption{\small The slice of the O'KKLT potential at vanishing axions $y$ and $\alpha$, multiplied by $10^{31}$, for the values of the parameters $A=1,~a=0.25,~W_0= -  10^{-12},~\mu^2=1.66\times 10^{-12},~L=10^{-3}$. The potential has a dS minimum at $\sigma\approx 123$, $x\sim 3 \cdot 10^{-7}$. The  gravitino mass in this example is $m_{3/2} \sim 600$ GeV. }\label{plot3uu} 
\end{figure}

Our result $x_{0}\approx{\sqrt 3  \over 6 }  \Lambda^2$ implies that the consistency condition ${S\bar S} =  {x^{2}} \ll { m^2\over \lambda^2 } $ is satisfied for
\be
m  \lesssim 10^{-2} \lambda^{3} \ .
\ee
The only parameters which are required for the calculations of the potential on the O'Raifeartaigh side  are   $\mu $ and $\Lambda$, whereas  $m= {\Lambda\sqrt c\over 4\pi}\lambda^{2}$, and $\lambda$ is a free parameter, which can be varied in order to satisfy the consistency conditions. One can easily find many sets of parameters which satisfy all required conditions. In particular, one may find the theory with the gravitino mass in the TeV range if one takes $\mu \sim 10^{{-6}}$, $\Lambda \sim 10^{{-3}}$, $m = 10^{{-6}}$, $\lambda \sim 10^{{-1}}$, see Figure \ref{plot3uu}. 

Thus the O'KKLT model can provide a consistent model of the F-term uplifting with the gravitino mass in the TeV range. On the other hand, if one attempts to describe superheavy gravitino, then it becomes difficult to satisfy all the consistency conditions, which require that some of the mass scales must be sufficiently small. In order to describe superheavy gravitino in such a model one may need to go beyond the simple approximations used to calculate the scalar potential and the  K\"ahler potential in our scenario.

\section{Light gravitino, vacuum stability and the KL model}

For many years, we wanted to have models with $m_{{3/2}}$ in the TeV range, and the O'KKLT model is well suited for this purpose. This fact is quite interesting, especially if one compares our model with the models with   the D-term uplifting, where it is very difficult to obtain a light gravitino   \cite{Burgess:2003ic}. In the O'KKLT model one can obtain a light gravitino, but it may be difficult to obtain a very heavy one. 

\begin{figure}[hbt]
\centering 
\includegraphics[scale=0.5]{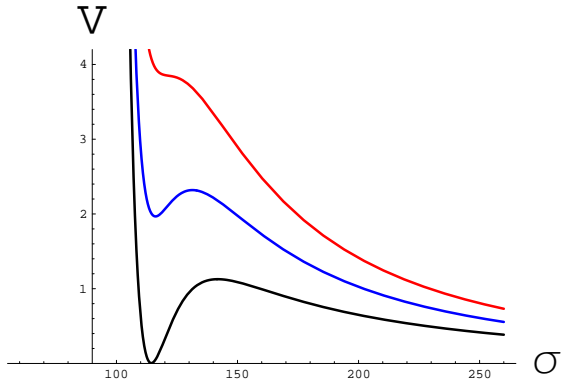} \caption{\small The lowest curve with dS minimum is the one from the uplifted KKLT model. The second one describes the  inflationary potential with the term $V_{\rm infl}={V(\phi)\over \sigma^3}$ added to the KKLT potential.  The top curve shows that when the  inflationary potential becomes too large, the barrier disappears, and the internal space decompactifies. This explains the constraint $H\lesssim   m_{3/2}$.}\label{plot1} 
\end{figure}

This property of our model may be desirable from the point of view of the phenomenological supergravity, but in the KKLT context it may lead to some cosmological problems   \cite{Kallosh:2004yh}. Indeed, the depth of the AdS vacuum in the KKLT scenario is given by $V_{\rm AdS} = -3 e^{K}|W|^2 = -3m^{2}_{{3/2}}$. Here $V_{\rm AdS}$ is the depth of the AdS minimum prior to the uplifting, and  $m_{{3/2}}$ is the gravitino mass after the uplifting. Uplifting creates the barrier separating the KKLT minimum from the 10D Minkowski Dine-Seiberg minimum. The height of the barrier is somewhat smaller than the depth of the original AdS minimum prior to the uplifting:
\be
V_{\rm barrier}\sim  |V_{AdS}| \approx 3 m_{3/2}^2  \ .
\ee
Inflation requires the existence of an additional contribution $V_{\rm infl}$ to the scalar potential, but all such contributions in the effective 4D theory have the following general structure: $V_{\rm infl} = {V(\phi)\over (\rho + \overline{\rho})^3}$, where $\phi$ is the inflaton field; see, for example,  the   second term in (\ref{potential}). Such terms destabilize the potential for $V_{\rm infl} \gg V_{\rm barrier} \sim m_{3/2}^2$, see Fig. \ref{plot1}. Since during inflation one has $V_{\rm infl} = H^{2}/3$, one finds the constraint on the Hubble constant during the last stage of inflation   \cite{Kallosh:2004yh}
\be
H \lesssim m_{{3/2}} \ .
\ee
If gravitino is heavy, $m_{3/2}\sim 10^{11}- 10^{16}$ GeV, the scale of inflation can be very high, and there is no destabilization of the volume modulus during inflation. However, if the mass of gravitino is light, e.g. $m_{3/2}\leq 1$  TeV, one would need to consider non-standard low-scale inflationary models. Such models do exist, but   still the constraint $H \lesssim m_{{3/2}}$ is quite restrictive and undesirable.

This problem was addressed in our paper  \cite{Kallosh:2004yh}, which we will call the KL model, to distinguish it from the simplest KKLT scenario.  We used the same   K\"ahler potential as in the KKLT model, but instead of the superpotential with one exponent we used the  racetrack-type superpotential with two exponents:
\be
K = - 3 \ln[(\rho + \overline{\rho})]\ , \qquad W = W_0 + Ae^{-a\rho}+ Be^{-b\rho} \ .
\label{KL}
\ee
The potential of this model can  describe either one or two AdS vacua, or, with extra fine tuning, one Minkowski vacuum and one AdS vacuum, both at a finite distance in the moduli space \cite{Kallosh:2004yh}. Here we will be interested in the case of two AdS vacua, such that $|V_{AdS_1}|\ll |V_{AdS_2}|$.

\begin{figure}[hbt]
\centering 
\includegraphics[scale=1]{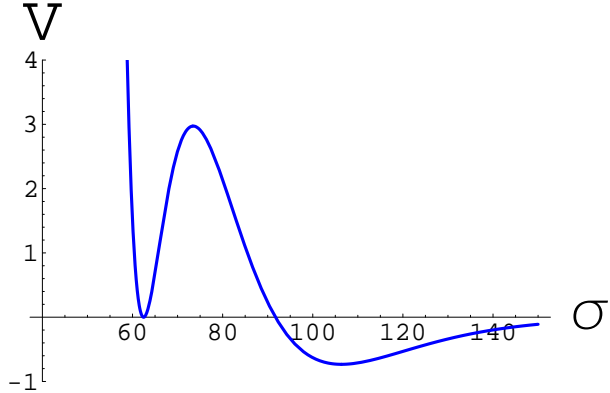} \caption{\small The KL potential multiplied by $10^{14}$, with the parameters $A=1,\ B=-1.03,\ a=2\pi/100,\ b=2\pi/99,\ W_0= - 2\times 10^{-4}$. The first minimum, corresponding to the supersymmetric Minkowski vacuum,   stabilizes the volume at $\sigma\approx 62$. If one slightly changes the parameters (e.g. takes $B = 1.032$), this minimum shifts down, and becomes a very shallow AdS minimum, see the thin black line in Fig. \ref{plot2}.}\label{plot2a} 
\end{figure}

Indeed, with one exponent  in the superpotential as in model (\ref{simple}) one cannot simultaneously solve both equations
$D_{\rho}W=0 $ and $W=0$, which is necessary 
to get a Minkowski minimum.\footnote{There is a no go theorem proposed in \cite{Brustein:2004xn,Gomez-Reino:2006dk,Lebedev:2006qq}, which states that it is impossible to have a dS/Minkowski minimum in the theory with the K\"ahler potential $K = - 3 \ln[\rho + \overline{\rho}]$, for any choice of the superpotential. The proof of this theorem  is correct  for  the supersymmetry breaking  Minkowski vacua with $D_{\rho}W\neq 0, W\neq 0$ \cite{Gomez-Reino:2006dk}, but it does not apply to the KL model, where  $D_{\rho}W=0, W=0$ in the  Minkowski minimum with unbroken supersymmetry   \cite{Kallosh:2004yh}.  It may be useful to remind here that the use of the supergravity function ${\cal G}= K+\ln |W|^2$ in studies of vacua with $W=0$ should be avoided, as explained in \cite{Kallosh:2000ve}, and such vacua should be studied separately. } 
However, with two exponents, as in model (\ref{KL}), this is possible. The solution requires the following relation between the parameters of the superpotential:
\be
-W_0=A\left |aA\over bB\right |^{a\over b-a}+A\left |aA\over bB\right |^{a\over b-a} \ .
\ee
The Minkowski minimum occurs at $\sigma={1\over a-b}\ln \left |aA\over bB\right |$; see Fig. \ref{plot2a}.

It has been observed in \cite{Ceresole:2006iq} that it is very easy (for example, by changing slightly the parameter B in the superpotential) to get models with extremely light gravitino and the AdS minimum replacing the exact Minkowski one with very small value of $|V_{AdS}|$ but large barrier separating this AdS from the next one and from the Minkowski vacuum at infinity. To make this model viable we need to uplift it to dS minimum by one of the available mechanisms. The $\overline{D3}$-brane uplifting  is possible. The D-term uplifting in a model with two exponents was not performed so far. It may require some effort to make it consistent with the gauge invariance of the superpotential. However,  the O'Raifeartaigh uplifting  works well for this model.  

The supergravity potential is based on                                                                                                                                                                                                                                                                                                                                                                                                              \be
W=W_0 + Ae^{-a\rho}+ B e^{-b\rho}- \mu^2 S, \qquad K= - 3 \ln(\rho + \overline{\rho})+ S\bar S - {(S\bar S)^2\over \Lambda^2} \ .
\label{model2}\ee
The complete potential $V(\sigma, \alpha, x, y)$ as a function of 4 scalars,
$
\rho= \sigma+i\alpha$ and   $S=x+iy
$,
is again easily computable in Mathematica, using  \cite{Kallosh:2004rs}, and, as before,    we are interested only in the region of small $S\bar S$. Investigation of this regime for the KL model practically coincides with the investigation for the O'KKLT model performed in the previous section. The axion fields vanish before and after the unification of the O'Raifeartaigh model with the KL model. The field $S$ after the uplifting takes  exactly the same value as in the KKLT scenario: $|S| = {\sqrt{3}\over 6}\Lambda^{2}$.

The influence of the O'Raifeartaigh uplifting of a shallow AdS minimum on the position of this minimum in this scenario is even smaller than the corresponding influence in the O'KKLT model. Indeed, since the required magnitude of the uplifting in this scenario is much smaller than the height of the barrier, all parameters of the  O'Raifeartaigh model can be taken many orders of magnitude smaller than the parameters  of the KL model. 

\begin{figure}[hbt]
\centering 
\includegraphics[scale=0.9]{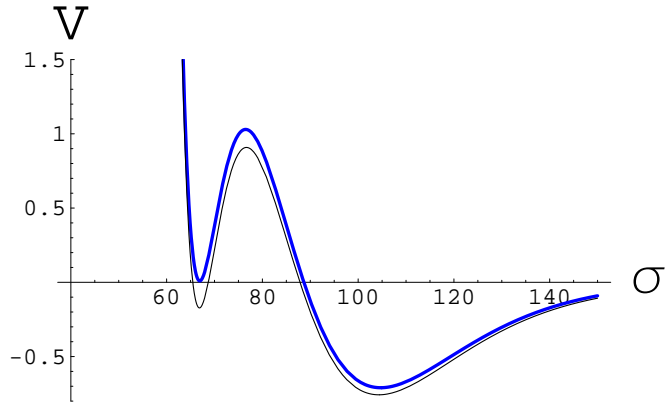} \caption{\small The thin black line shows the potential in the KL model, multiplied by $10^{14}$, for the values of the parameters $A=1,\ B=-1.032,\ a=2\pi/100,\ b=2\pi/99,\ W_0= - 2\times 10^{-4}$. The shallow AdS minimum (almost Minkowski)   stabilizes the volume at $\sigma\approx 67$. The thick blue line shows the potential after the  O'Raifeartaigh  uplifting with   $ \mu^2=0.66 \times 10^{-4}$, $L=10^{-3}$. The AdS minimum after the uplifting  becomes a (nearly Minkowski) dS minimum.}\label{plot2} 
\end{figure}

An example of the KL potential before and after the uplifting (thin and thick lines) is shown in Fig. \ref{plot2}.  The depth of the shallow AdS minimum prior to the uplifting (thin line) corresponds to $3m^{2}_{3/2}$. This depth, and the required magnitude of uplifting, controlled by  the parameter $\mu^{2}$, can be made arbitrarily small by a slight change of the parameter $B$, which practically does not affect the height of the barrier. Therefore in this scenario the barrier can be many orders higher than $m^{2}_{3/2}$. In this figure we have made a relatively large modification of $B$, which leads to the large gravitino mass, but we did it only for the purpose of making the modification of the potential visible.

The main new feature of the KL model as compared with the O'KKLT model is that one can fine-tune the gravitino mass squared to be extremely small as compared to the height of the barrier. This allows inflation with $H \gg m_{3/2}$ \cite{Kallosh:2004yh}.

\section{Discussion}

The existence of a tiny positive cosmological constant makes it quite important to stabilize all moduli in a dS state with a positive vacuum energy. In this note we  proposed to combine two simple models to achieve this purpose. The KKLT model originating from string theory
brings in the idea of $\sim 10^{500}$ various vacua, mostly supersymmetric AdS  vacua with negative energy. Stabilization of closed string theory moduli is due to non-perturbative effects, like gaugino condensation or string instantons. The second ingredient of the proposed unified model is a generic model of dynamical supersymmetry breaking where  non-perturbative corrections play an important role in stabilization of moduli (these moduli in string theory may come from the open string sector). A typical representative of such models is the   O'Raifeartaigh model with the Coleman-Weinberg quantum corrections. Moduli stabilization in this model occurs near the origin of the moduli space and results in the existence of a dS vacuum with positive energy. When these two models are unified, they affect each other in a minor way in all respects but one: the negative AdS energy of the KKLT model is nearly compensated by the positive dS energy of the O'Raifeartaigh model, which leads to a nearly Minkowski space which we observe.

In the paper we present the detailed description of the unification of KKLT model with quantum corrected O'Raifeartaigh model, which we called O'KKLT. The effect which each of these two models has on the other one is computable. In particular, one can find a class of the O'KKLT models with light gravitinos,   and achieve vacuum stability  during the cosmological evolution.

\

\noindent{\large{\bf Acknowledgments}}

\noindent
It is a pleasure to thank S. Dimopoulos, M. Dine, B. Freivogel, R. Kitano, L. Susskind, 
J.~Wacker and E. Witten for stimulating conversations.   This work  was supported by NSF grant PHY-0244728.

\vskip 1cm

\providecommand{\href}[2]{#2}
\begingroup\begin{thebibliography}{99}

\bibitem{O'Raifeartaigh:1975pr}
  L.~O'Raifeartaigh,
  ``Spontaneous Symmetry Breaking For Chiral Scalar Superfields,''
  Nucl.\ Phys.\ B {\bf 96}, 331 (1975).
  
\bibitem{Huq:1975ue}
  M.~Huq,
  ``On Spontaneous Breakdown Of Fermion Number Conservation And
Supersymmetry,''
  Phys.\ Rev.\ D {\bf 14}, 3548 (1976).
  
\bibitem{Witten:1981kv}
  E.~Witten,
  ``Mass Hierarchies In Supersymmetric Theories,''
  Phys.\ Lett.\ B {\bf 105}, 267 (1981).

\bibitem{Poppitz:1998vd}
  E.~Poppitz and S.~P.~Trivedi,
  ``Dynamical supersymmetry breaking,''
  Ann.\ Rev.\ Nucl.\ Part.\ Sci.\  {\bf 48}, 307 (1998)
  [arXiv:hep-th/9803107].




  
\bibitem{Chacko:1998si}
  Z.~Chacko, M.~A.~Luty and E.~Ponton,
  ``Calculable dynamical supersymmetry breaking on deformed moduli spaces,''
  JHEP {\bf 9812}, 016 (1998)
  [arXiv:hep-th/9810253].
  
    
\bibitem{Intriligator:2006dd}
  K.~Intriligator, N.~Seiberg and D.~Shih,
  ``Dynamical SUSY breaking in meta-stable vacua,''
  JHEP {\bf 0604}, 021 (2006)
  [arXiv:hep-th/0602239].

\bibitem{Dine:2006ii}
  M.~Dine, R.~Kitano, A.~Morisse and Y.~Shirman,
  ``Moduli decays and gravitinos,''
  Phys.\ Rev.\ D {\bf 73}, 123518 (2006)
  [arXiv:hep-ph/0604140];
  R.~Kitano,
 ``Dynamical GUT breaking and mu-term driven supersymmetry breaking,''
  arXiv:hep-ph/0606129.
  
\bibitem{Kitano:2006wz}
  R.~Kitano,
  ``Gravitational gauge mediation,''
  Phys.\ Lett.\ B {\bf 641}, 203 (2006)
  [arXiv:hep-ph/0607090].



\bibitem{Craig:2006kx}
  N.~J.~Craig, P.~J.~Fox and J.~G.~Wacker,
  ``Reheating metastable O'Raifeartaigh models,''
  arXiv:hep-th/0611006.
  
    
\bibitem{Kachru:2003aw}
  S.~Kachru, R.~Kallosh, A.~Linde and S.~P.~Trivedi,
``De Sitter vacua in string theory,''
  Phys.\ Rev.\ D {\bf 68}, 046005 (2003)
  [\href{http://arXiv.org/abs/hep-th/0301240}{{\tt hep-th/0301240}}].
  
    
\bibitem{Burgess:2003ic}
  C.~P.~Burgess, R.~Kallosh and F.~Quevedo,
  ``de Sitter string vacua from supersymmetric D-terms,''
  JHEP {\bf 0310}, 056 (2003)
  [arXiv:hep-th/0309187];
  G.~Villadoro and F.~Zwirner,
  ``de Sitter vacua via consistent D-terms,''
  Phys.\ Rev.\ Lett.\  {\bf 95}, 231602 (2005)
  [arXiv:hep-th/0508167];
  A.~Achucarro, B.~de Carlos, J.~A.~Casas and L.~Doplicher,
  ``de Sitter vacua from uplifting D-terms in effective supergravities from
  realistic strings,''
  JHEP {\bf 0606}, 014 (2006)
  [arXiv:hep-th/0601190];
   K.~Choi and K.~S.~Jeong,
  ``Supersymmetry breaking and moduli stabilization with anomalous U(1) gauge
  symmetry,''
  JHEP {\bf 0608}, 007 (2006)
  [arXiv:hep-th/0605108]; E.~Dudas and Y.~Mambrini,
  ``Moduli stabilization with positive vacuum energy,''
  JHEP {\bf 0610}, 044 (2006)
  [arXiv:hep-th/0607077];
  M.~Haack, D.~Krefl, D.~Lust, A.~Van Proeyen and M.~Zagermann,
  ``Gaugino condensates and D-terms from D7-branes,''
  arXiv:hep-th/0609211;
  C.~P.~Burgess, J.~M.~Cline, K.~Dasgupta and H.~Firouzjahi,
  ``Uplifting and inflation with D3 branes,''
  arXiv:hep-th/0610320.


  
 
\bibitem{Saltman:2004sn}
  A.~Saltman and E.~Silverstein,
  ``The scaling of the no-scale potential and de Sitter model building,''
  JHEP {\bf 0411}, 066 (2004)
  [arXiv:hep-th/0402135].
  
\bibitem{Gomez-Reino:2006dk}
  M.~Gomez-Reino and C.~A.~Scrucca,
``Locally stable non-supersymmetric Minkowski vacua in supergravity,''
  JHEP {\bf 0605}, 015 (2006)
  [arXiv:hep-th/0602246].



\bibitem{Lebedev:2006qq}
  O.~Lebedev, H.~P.~Nilles and M.~Ratz,
 ``de Sitter vacua from matter superpotentials,''
  Phys.\ Lett.\ B {\bf 636}, 126 (2006)
  [arXiv:hep-th/0603047].

\bibitem{Brummer:2006dg}
  F.~Brummer, A.~Hebecker and M.~Trapletti,
``SUSY breaking mediation by throat fields,''
  Nucl.\ Phys.\ B {\bf 755}, 186 (2006)
  [arXiv:hep-th/0605232].

\bibitem{Dudas:2006gr}
  E.~Dudas, C.~Papineau and S.~Pokorski,
 ``Moduli stabilization and uplifting with dynamically generated F-terms,''
  arXiv:hep-th/0610297.
  
\bibitem{Abe:2006xp}
  H.~Abe, T.~Higaki, T.~Kobayashi and Y.~Omura,
  ``Moduli stabilization, F-term uplifting and soft supersymmetry breaking
terms,''
  arXiv:hep-th/0611024.
  
\bibitem{Dine:2004is}
  M.~Dine, E.~Gorbatov and S.~D.~Thomas,
``Low energy supersymmetry from the landscape,''
  arXiv:hep-th/0407043;
  M.~Dine,
``The intermediate scale branch of the landscape,''
  JHEP {\bf 0601}, 162 (2006)
  [arXiv:hep-th/0505202].
  
\bibitem{Westphal:2005yz}
  V.~Balasubramanian and P.~Berglund,
``Stringy corrections to Kaehler potentials, SUSY breaking, and the
cosmological constant problem,''
  JHEP {\bf 0411}, 085 (2004)
  [arXiv:hep-th/0408054];
  A.~Westphal,
 ``Eternal inflation with alpha' corrections,''
  JCAP {\bf 0511}, 003 (2005)
  [arXiv:hep-th/0507079];
  A.~Westphal,``de Sitter String Vacua from K\"{a}hler Uplifting'', work in progress.
  
  
\bibitem{Izawa:1996pk}
  K.~I.~Izawa and T.~Yanagida,
 ``Dynamical Supersymmetry Breaking in Vector-like Gauge Theories,''
  Prog.\ Theor.\ Phys.\  {\bf 95}, 829 (1996)
  [arXiv:hep-th/9602180];
  K.~A.~Intriligator and S.~D.~Thomas,
  ``Dynamical Supersymmetry Breaking on Quantum Moduli Spaces,''
  Nucl.\ Phys.\ B {\bf 473}, 121 (1996)
  [arXiv:hep-th/9603158].
  
\bibitem{Dimopoulos:1997ww}
  S.~Dimopoulos, G.~R.~Dvali, R.~Rattazzi and G.~F.~Giudice,
  ``Dynamical soft terms with unbroken supersymmetry,''
  Nucl.\ Phys.\ B {\bf 510}, 12 (1998)
  [arXiv:hep-ph/9705307].
  
\bibitem{Kallosh:2004yh}
  R.~Kallosh and A.~Linde,
``Landscape, the scale of SUSY breaking, and inflation,''
  JHEP {\bf 0412}, 004 (2004)
  [arXiv:hep-th/0411011];
J.~J.~Blanco-Pillado, R.~Kallosh and A.~Linde,
``Supersymmetry and stability of flux vacua,'' JHEP {\bf 0605}, 053 (2006)
  [arXiv:hep-th/0511042].

  
  
\bibitem{Kallosh:2004rs}
  R.~Kallosh and S.~Prokushkin,
  ``Supercosmology,''
  arXiv:hep-th/0403060.


\bibitem{Polonyi:1977pj}
J.~Pol\'{o}nyi,
``Generalization of the massive scalar multiplet coupling to the supergravity,''
Hungary Central Inst Res - KFKI-77-93 (July 78) 5p.



\bibitem{Brustein:2004xn}
  R.~Brustein and S.~P.~de Alwis,
``Moduli potentials in string compactifications with fluxes: Mapping the
discretuum,''
  Phys.\ Rev.\ D {\bf 69}, 126006 (2004)
  [arXiv:hep-th/0402088].
 

\bibitem{Kallosh:2000ve}
  R.~Kallosh, L.~Kofman, A.~D.~Linde and A.~Van Proeyen,
 ``Superconformal symmetry, supergravity and cosmology,''
  Class.\ Quant.\ Grav.\  {\bf 17}, 4269 (2000)
  [Erratum-ibid.\  {\bf 21}, 5017 (2004)]
  [arXiv:hep-th/0006179].










  
\bibitem{Ceresole:2006iq}
  A.~Ceresole, G.~Dall'Agata, A.~Giryavets, R.~Kallosh and A.~Linde,
  ``Domain walls, near-BPS bubbles, and probabilities in the landscape,''
  Phys.\ Rev.\ D {\bf 74}, 086010 (2006)
  [arXiv:hep-th/0605266].



 


     
\end{thebibliography}\endgroup
\end{document}